\documentclass[10pt]{iopart}
\usepackage{graphics}
\usepackage{subfig}
\usepackage{hyperref}
\expandafter\let\csname equation*\endcsname\relax
\expandafter\let\csname endequation*\endcsname\relax
\usepackage{amsmath}
\usepackage{graphicx}
\usepackage{cite}
\usepackage[utf8]{inputenc}
\DeclareGraphicsExtensions{.jpg,.pdf,.mps,.png,.bmp}

\begin{document}

\title[Ince-Gauss Photons in Turbulent Atmosphere]{Ince-Gauss Photons in Turbulent Atmosphere: Effect of quantum numbers on beam resilience}
\author{Emmanuel Narváez Castañeda$^{1,2}$, Roberto Ramírez Alarcón$^{2,*}$, José César Guerra Vázquez $^{1,2}$, Imad Agha$^{1}$,  Qiwen Zhan$^{3}$, and William N. Plick$^{1,**}$}

\address{$^{1}$ Department of Electro-Optics and Photonics, University of Dayton, Dayton, Ohio 45469, USA}
\address{$^{2}$ Centro de Investigaciones en Óptica A.C., Loma del Bosque 115, Colonia Lomas del Campestre, 37150
León Guanajuato, México}
\address{$^{3}$ School of Optical-Electrical and Computer Engineering, University of Shanghai for Science and
Technology, Shanghai 200093, China}
\ead{*roberto.ramirez@cio.mx}
\ead{**wplick1@udayton.edu}

\begin{abstract}
In this work, we present an extensive analysis on the nature and performance of Ince-Gauss beams, elliptical solutions of the paraxial wave equation that have orbital angular momentum, as information carriers in turbulent atmosphere. We perform numerical simulations of the propagation of these beams, and focus on the effects that the order, degree and ellipticity parameters have on the robustness of the beams. We find that the choice of basis in which a mode is constructed does not greatly influence the mode performance and it is instead strongly affected by the combination of order and degree values.
\end{abstract}
\maketitle
\ioptwocol
%
%
%
%
%

\section{Introduction}
 Today's state of the art in optical communications consists mainly of fiber optics based systems, but while such systems are very mature in development, they represent a high cost if they are to be considered to fully replace legacy wire communication networks. This is why they are not, in some cases, a viable solution for last mile connectivity or for the building of networks in remote places \cite{first}. For these reasons and others, like the possibility to use different basis like polarization and optical angular momentum of light to encode information that is resistant to eavesdropping, free space optical (FSO) communication systems represent an excellent alternative \cite{Third,second, fourth, fifth, Gibson:04}.\\
The vast majority of studies on FSO systems focus on the properties of fundamental Gaussian beams. However, in recent years other kinds of modified Gaussian beams, structured light beams, have been studied, as it has been found that initial beam properties such as shape, phase, size, coherence, etc. strongly affect their performance \cite{tenth}. Of special interest are structured light beams that carry orbital angular momentum (OAM) \cite{second, eleventh, twelve}, which began to be broadly studied following the work by Allen et al. \cite{Allen}.\\

OAM is an intrinsic property of light that resides in a theoretically infinite-dimensional space, which is spanned by an infinite-dimensional basis of orthogonal modes, each with the capacity of being an independent channel of information. Although in an actual FSO system the finite number of OAM channels is determined by the space-bandwidth, comparing this to the capacity of another usually used degree of freedom such as polarization, which resides in a two-dimensional space, the usage of orbital angular momentum as information carrier results in a massive advantage due the raw number of communication channels \cite{catorce}. The importance of OAM carrying beams resides as well in a broad field of applications and studies such as: optical tweezers, imaging and microscopy, biomedicine, metrology, astronomy, fluids mechanics \cite{diezsiete,diezocho, dieznueve,veinte, veintiuno, veintidos, veintitres} and even in tests and applications of quantum mechanics, such as quantum entanglement and high dimensional quantum key distribution \cite{Zeferino, veinticinco}.\\ 

While the use of optical vortex beams for free space applications has been studied in free space quantum key distribution and satellite communications systems \cite{veintiseis,veintisiete}, most of the work has been done using Laguerre-Gaussian (LG) beam profiles, natural solutions of the paraxial wave equation in cylindrical coordinates that carry orbital angular momentum. However, limited study has been done using a more general beam profile family of modes that also contains orbital angular momentum: The Ince-Gaussian beams, elliptical solutions of the paraxial wave equation that have an additional degree of freedom in their ellipticity parameter and that were introduced first by Bandres et al. \cite{bandres1, bandres2}. This characteristic makes them potential candidates for usage in applications involving optical vortex beams, particularly those regarding free space propagation. Besides few studies on FSO communication systems for these beams \cite{Eyyuboglu, Gu, Zhu:20}, Ince-Gauss (IG) beams have been used in applications such as OAM entanglement \cite{Plick, KRENNPLICK}, optical tweezers \cite{tweezers} and have exhibited lots of intriguing phenomenon when propagating in different types of media like strongly nonlocal nonlinear media \cite{nonlocal}, elliptical core few-mode fibers \cite{fiber}, or even interacting with atoms \cite{Yu:21}.\\

In this work, we investigate properties of Ince-Gaussian beams by presenting a procedure on how the propagation of these modes through atmospheric turbulence can be simulated. A quantitative and qualitative analysis on the performance and behavior of these beams is made with special attention to their usage in FSO communication systems. In the second section, we present a brief mathematical description of Ince-Gauss beams and their properties. In the third one, we show the numerical modeling method for propagation simulation of these beams. In the fourth section we expose and analyze the outcome from said simulations by varying the beam parameters, and finally in the fifth section we give conclusions and perspectives about the results.

\section{Ince-Gaussian beams}

Here we provide an introduction to the origin and properties of elliptical vortex beams for convenience of the reader. This information can be bypassed by those with strong familiarity of the subject.\\

The \textit{Paraxial Wave Equation (PWE)} is derived from the electromagnetic wave equation by assuming that the amplitude of the beam is a slowly varying function with the propagation axis \cite{Svelto}:

\begin{equation}
     \nabla^{2}_{t}\Phi(\vec{r}) - 2ik\frac{\partial \Phi(\vec{r})}{\partial z}=0\textrm{,}
     \label{pwe}
 \end{equation}

\noindent
where $\nabla^{2}_{t} = \partial^2/\partial x^2 + \partial^2/\partial y^2
$ is the transverse Laplacian for the $x$ and $y$ directions, $\Phi(\vec{r})$ is the scalar beam as function of position and $k$ is the wave number.\\

The lowest order solution of the \textit{PWE} is a gaussian beam, described by \cite{partha}:

 \begin{equation}
     \Phi_{G}(\vec{r}) = \frac{w_{0}}{w(z)}\textrm{exp}\left[- i k z-\frac{r^{2}}{w^{2}(z)}-i\frac{kr^{2}}{2R(z)}+i\phi_{g}(z)\right]\textrm{,}
     \label{GaussianBeamEqn}
 \end{equation}
 
 \noindent
  where $r$ is the radial distance from the center axis of the beam, $w_{0}$ corresponds to the beam width at $z=0$, $w(z)$ describes the width of the beam, $R(z)$ is the radius of curvature of the wavefront and $\phi_{g}(z)$ is the Gouy phase shift.\\
  
   Although equation (\ref{GaussianBeamEqn}) is the most straightforward solution to the \textit{PWE}, other solutions can be derived if we assume that the beam profile is the aforementioned gaussian beam solution \textit{modulated} by some other kind of transverse profile:
 
 \begin{equation}
     \Phi(\vec{r}) = A_{t}(\vec{r_{t}})[iZ(z)]\Phi_{G}(\vec{r})\textrm{,}
     \label{gneqn}
 \end{equation}
 
 \noindent
 where $A_{t}(\vec{r_{t}})$ is a function that depends only on the transverse coordinates of the beam contained within the transverse position vector $\vec{r_{t}}$. In the case of choosing elliptical coordinates and assuming a $z=0$ propagation distance, the solution of the \textit{PWE} yields the Ince-Gauss (IG) beams \cite{INCEEQ,bandres1,bandres2}:
 
 \begin{multline}
    IG^{e,\varepsilon}_{p,m}=\frac{\textit{C}w_{0}}{w(z)}C_{p}^{m}(i\xi,\varepsilon)C_{p}^{m}(\eta,\varepsilon)\textrm{exp}\left[\frac{-r^{2}}{w^{2}(z)}\right]\\\textrm{exp}\left[i\left(kz+\frac{k r^{2}}{2R(z)}-(p+1)\phi_{g}(z)\right)\right]\textrm{,}
    \label{IGeqn1}
    \end{multline}
    
\begin{multline}
    IG^{o,\varepsilon}_{p,m}=\frac{\textit{S} w_{0}}{w(z)}S_{p}^{m}(i\xi,\varepsilon)S_{p}^{m}(\eta,\varepsilon)\textrm{exp}\left[\frac{-r^{2}}{w^{2}(z)}\right]\\\textrm{exp}\left[i\left(kz+\frac{k r^{2}}{2R(z)}-(p+1)\phi_{g}(z)\right)\right]\textrm{,}
    \label{IGeqn2}
\end{multline}

\noindent
here, $z=z$ and the elliptical coordinates $\xi$ and $\eta$ are defined as $x = f(z)\mathrm{cosh(\xi)cos(\eta)}$ and $y = f(z)\mathrm{sinh(\xi)sin(\eta)}$, with $\xi\in [0,\infty)$ and $\eta\in[0,2\pi)$, being $\xi$ and $\eta$ the radial and angular elliptic variables. $f(z) = f_{0}w(z)/w_{0}$ is the semi focal separation with $f_{0}$ the initial semi focal separation and $e$ and $o$ refer to the parity of the beam, \textit{even} or \textit{odd}. The suffix $p$ corresponds to the order and $m$ to the degree of the modes, both having the same parity integer numbers and with the condition $p\geq m\geq 0$ for even numbers and $p\geq m\geq 1$ for odd numbers. $C_{m}^{p}$ and $S_{m}^{p}$ are the even and odd Ince-Polynomials of order $p$ and degree $m$ and \textit{C} and \textit{S} are normalization constants \cite{bandres2}. For these beams, The $\varepsilon$ parameters adjust the ellipticity of the transverse structure, while the parameters $w_{0}$ and $f_{0}$ scale the physical size of the mode.\\

\begin{figure*}[h]
	\begin{center}
	\subfloat[\normalsize{$LG_{1,3}^{o}$ = $IG_{5,3}^{o,0}$} ]{\includegraphics[scale = 0.11]{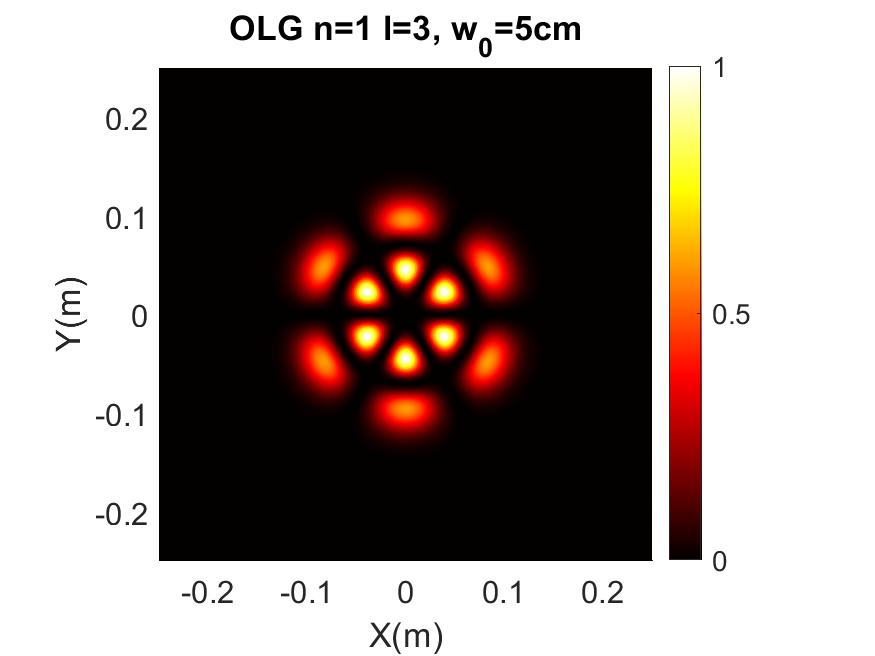}}
    \subfloat[\normalsize{$IG_{5,3}^{o,1}$}]{\includegraphics[scale = 0.11]{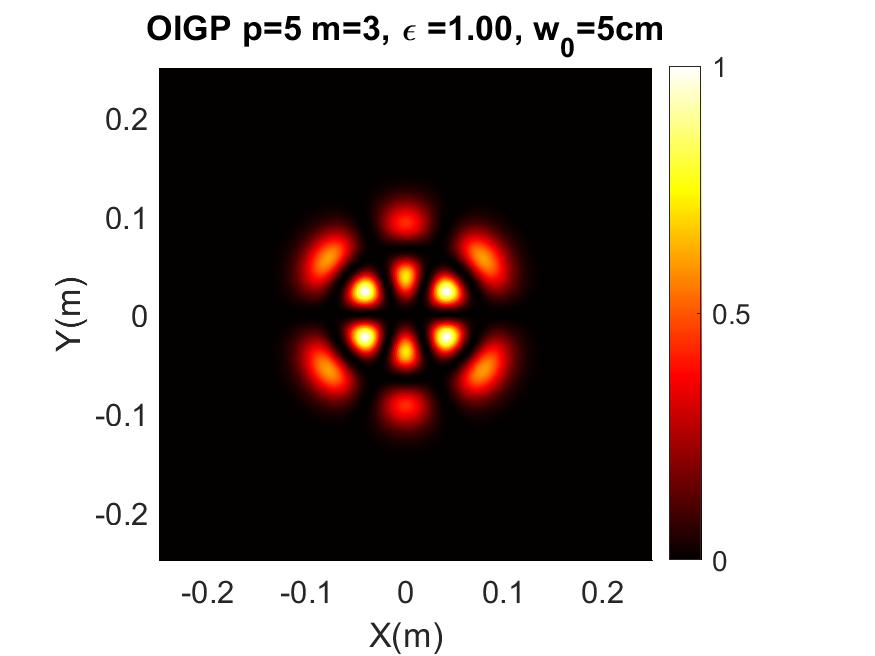}}
    \subfloat[\normalsize{$IG_{5,3}^{o,4.2}$}]{\includegraphics[scale = 0.11]{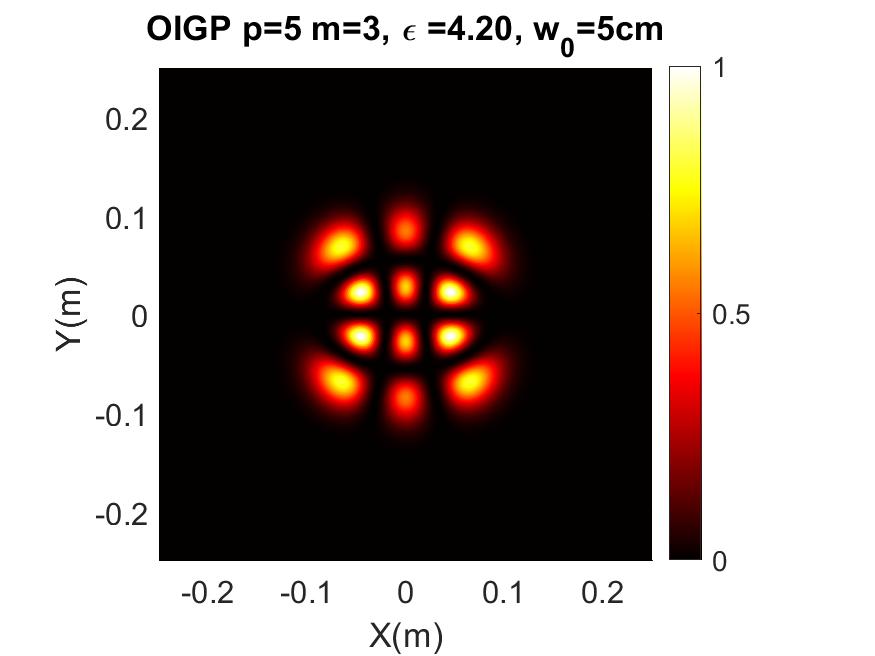}}
    \subfloat[\normalsize{$HG_{2,3}^{o}$ = $IG_{5,3}^{o,\infty}$}]{\includegraphics[scale = 0.11]{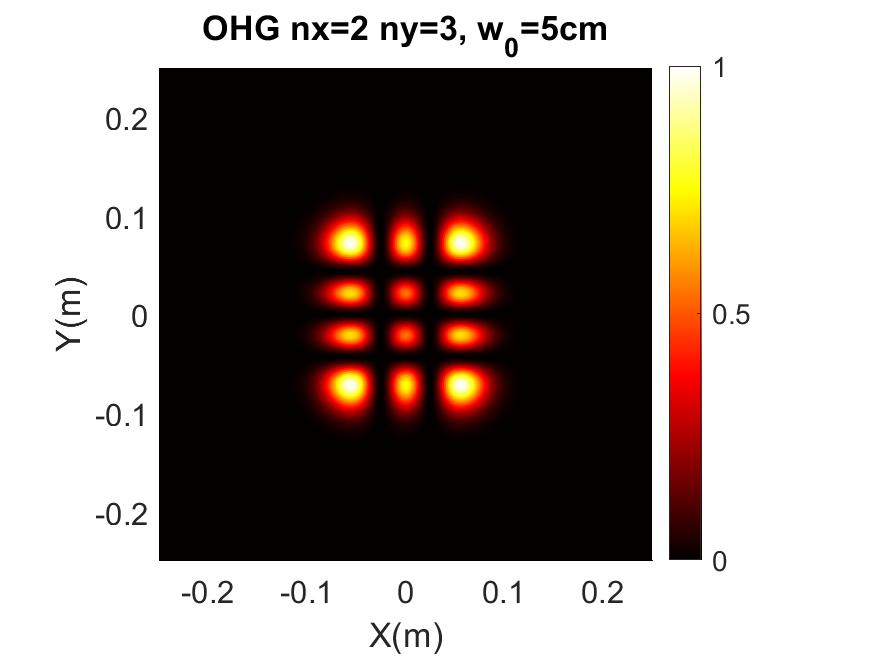}}
	\caption{Evolution of a $IG_{5,3}^{o}$ beam, $w_{0}=5cm$, by varying the ellipticity parameter from $\varepsilon=0$ to $\varepsilon=\infty$.}
	\label{oddevolution}
	\end{center}
\end{figure*}

Ince-Gauss modes are an orthogonal basis of solutions to the \textit{PWE}, and thus, any superposition of IG beams can be described in any other basis of solutions of the \textit{PWE}, like Hermite-Gaussian (\textit{HG}) and Laguerre-Gaussian (\textit{LG}). The transition of a IG mode into a LG mode occurs when $\varepsilon = 0$ (circular cylindrical coordinates), while the transition into HG when $\varepsilon = \infty$ (Cartesian coordinates). Similar to how $\varepsilon = 0$ and $\varepsilon=\infty$ define a complete orthogonal basis of modes, any value of the ellipticity parameter describes an individual orthogonal basis \cite{bandres2}. The transition of the indices, for LG modes is related as: $m=l$ and $p= 2n+l$, while for HG these relations depend on the parity of the IG mode: For even ones $n_{x} = m$ and $n_{y} = p-m$, while for odd ones $n_{x} = m-1$ and $n_{y} = p-m+1$, where $l$ and $n$ are the azimuthal and radial parameters of LG modes and $n_{x}$ and $n_{y}$ are the \textit{x-y} parameters of the HG modes. These relations of indices come from the fact that the Gouy shifts of the beams should match and are modulated by the indices \cite{bandres2}. The transition of orthogonal basis of a mode due to the ellipticity parameter is depicted in Figure \ref{oddevolution}, where the evolution of a $IG_{5,3}^{o,\varepsilon}$ with the value of $\varepsilon$ is shown, transforming the original $LG_{1,3}^{o}$ mode into a $HG_{2,3}$ mode for extreme values of $\varepsilon$. The symmetry of the beam is transformed due to the coordinate system used in each beam, going from azimuthal to rectangular.\\

It is a common misconception with this formalism that since classical electromagnetic theory has been used in the derivation of these equations, it is a classical result that does not hold in the quantum domain, and that the above equations represent ``beams'' and not ``photons''. This point of view is however incorrect. This can be seen by examining the paraxial wave equation, Eq.(\ref{pwe}). Making the substitution of time for $z$-position we find that the PWE is equivalent to the time-dependent Schr\"{o}dinger equation in free space. Since propagation distance and time are equivalent for light this substitution is well-formed. Thus results formulated as superpositions of electromagnetic fields obeying the PWE hold just as well for individual photons in quantum superpositions of the equivalent eigenmodes. Briefly, the quantum and ``beam-like'' descriptions of light fields are for \emph{most} purposes equivalent (at minimum they are the same for the work presented here).  \\

A particular characteristic of solutions of the \textit{PWE} is that certain superpositions of LG, HG and IG modes give place to \textit{helical modes}, which at localized parts of their phase profiles have polar phase indeterminations characterized by discontinuities with value $\pm 2\pi l'$,  $l'\in Z$. These result in local zeros of intensity in the beam profile and an orbital angular momentum \textit{(OAM)} value of $l'\hbar$ per photon. The value $l$ is also known as the \textit{topological charge} of the beam. Helical IG modes result from the superposition:

\begin{equation}
    HIG_{p,m}^{\pm,\varepsilon}(\xi,\eta,\varepsilon) = \frac{1}{\sqrt{2}}\left[IG_{p,m}^{e, \varepsilon}(\xi,\eta,\varepsilon)\pm iIG_{p,m}^{o,\varepsilon}(\xi,\eta,\varepsilon)\right]\textrm{.}
    \label{HIG}
\end{equation}

\begin{figure}
	\begin{center}
	\subfloat[\normalsize{$HLG_{1,3}$\\ Intensity}]{\includegraphics[scale = 0.11]{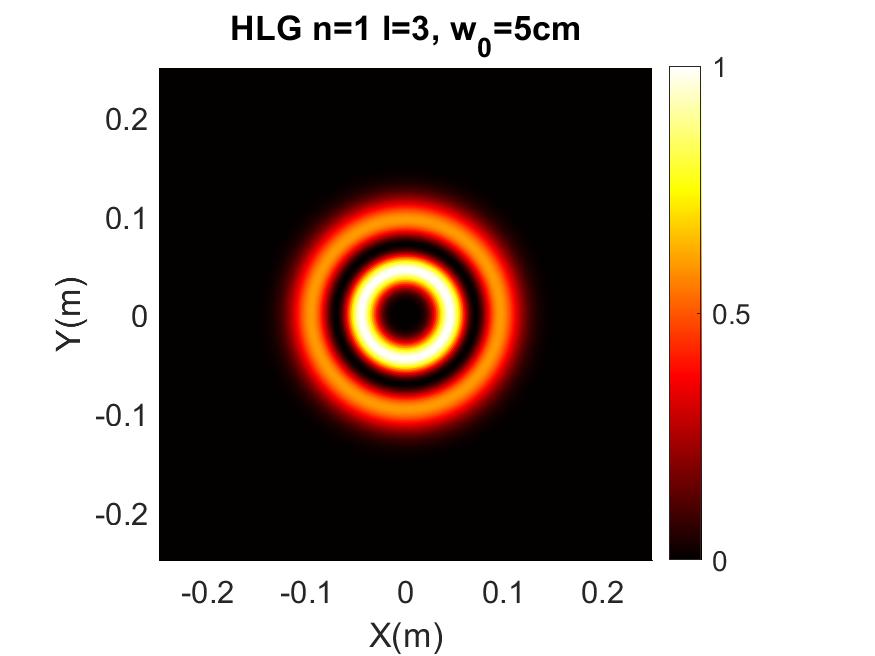}}
	\subfloat[\normalsize{$HIG_{5,3}^{1}$\\ Intensity}]{\includegraphics[scale = 0.11]{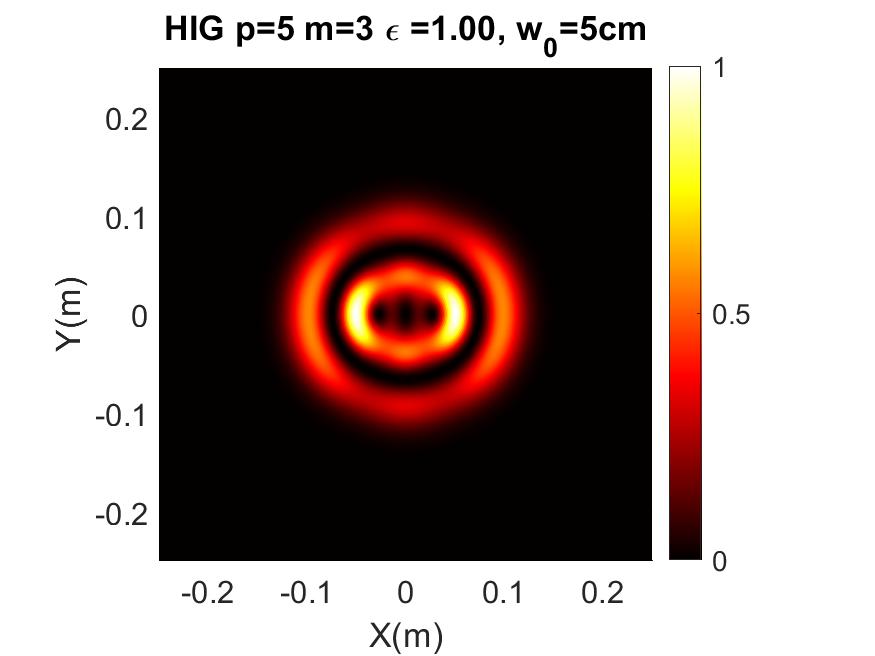}}\\
	\subfloat[\normalsize{$HIG_{5,3}^{4.2}$\\ Intensity}]{\includegraphics[scale = 0.11]{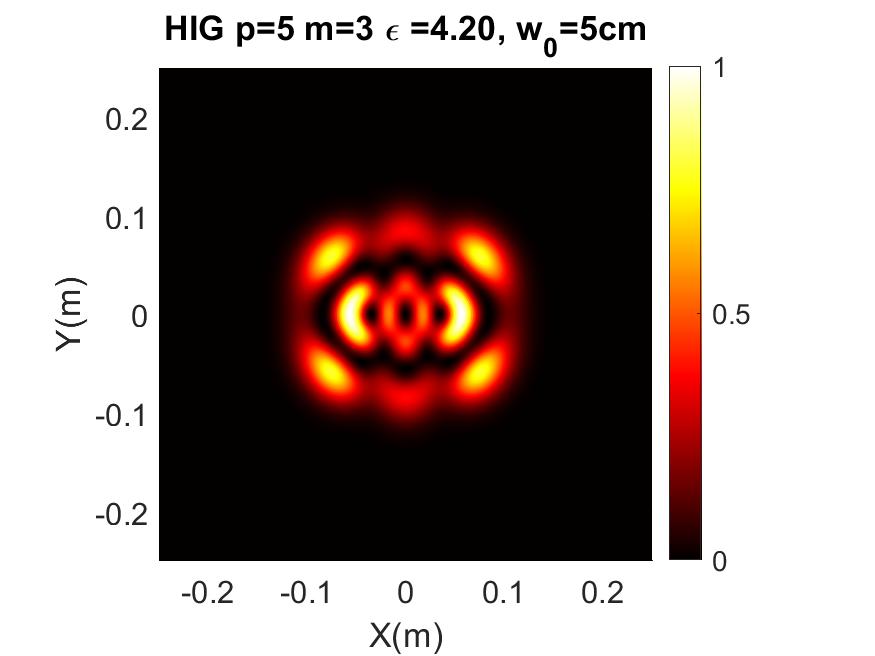}}
	\subfloat[\normalsize{$HHG_{2,3}$\\ Intensity}]{\includegraphics[scale = 0.11]{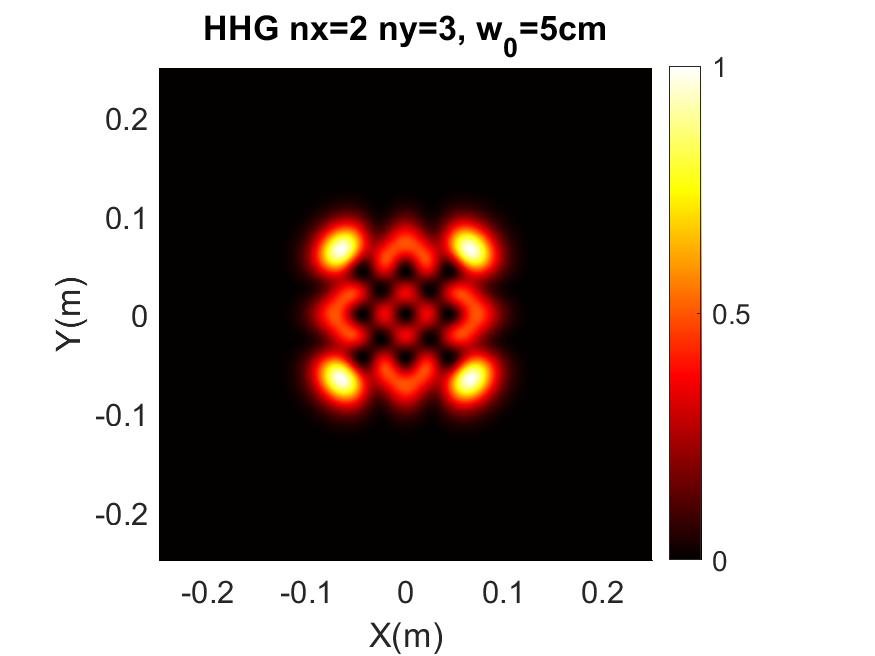}}\\
	\subfloat[\normalsize{$HLG_{1,3}$ Phase}]{\includegraphics[scale = 0.11]{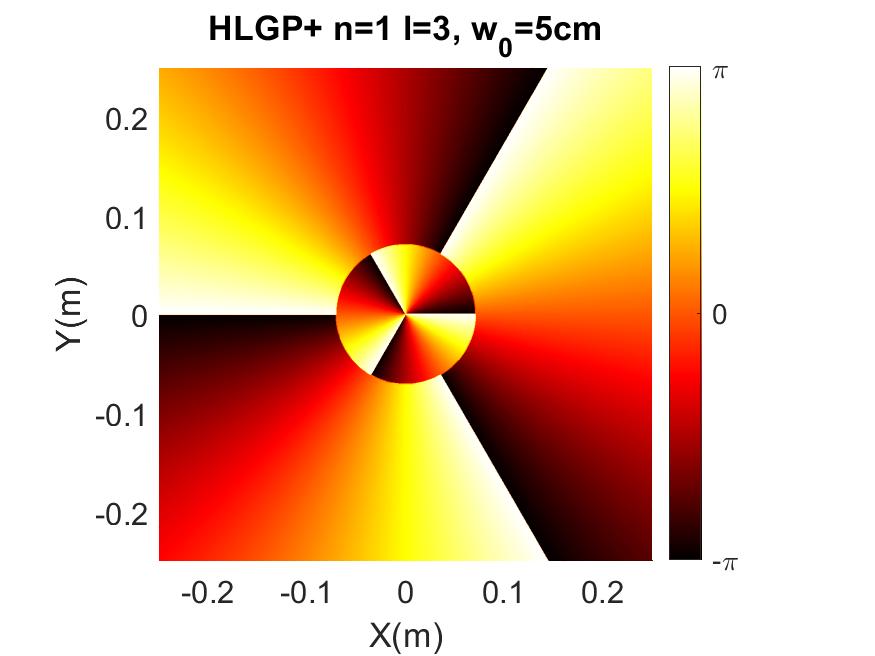}}
	\subfloat[\normalsize{$HIG_{5,3}^{1}$ Phase}]{\includegraphics[scale = 0.11]{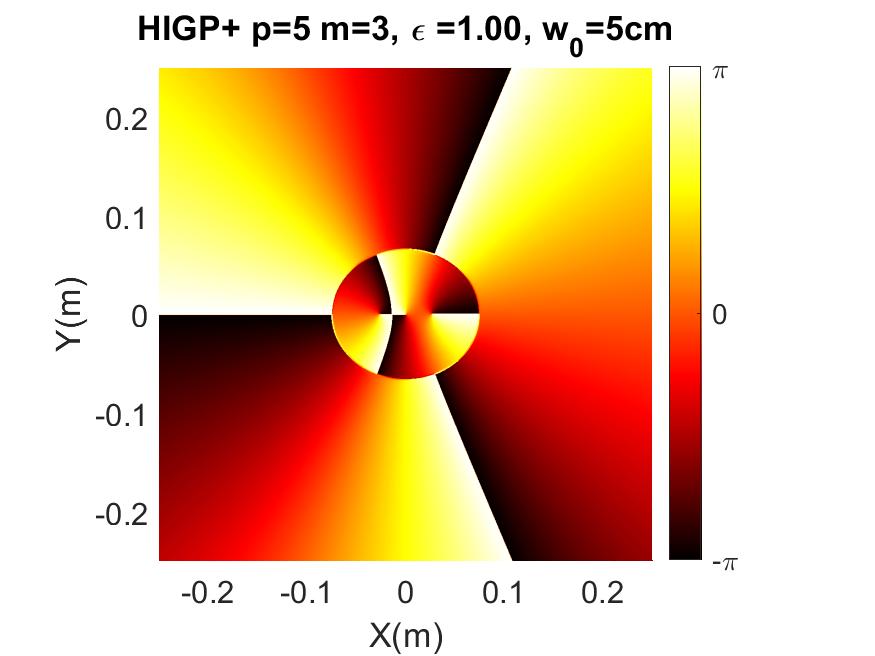}}\\
	\subfloat[\normalsize{$HIG_{5,3}^{4.2}$ Phase}]{\includegraphics[scale = 0.11]{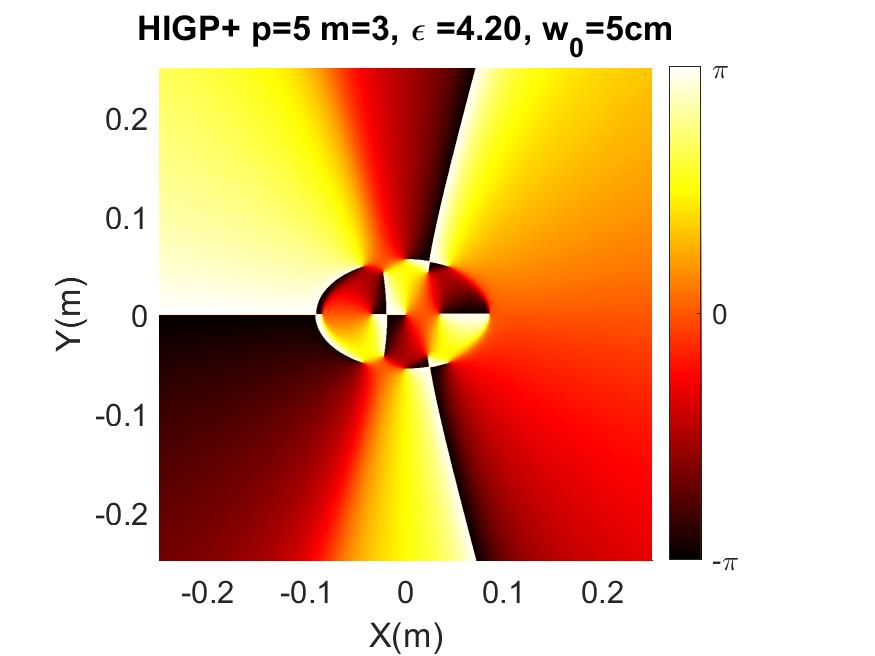}}
	\subfloat[\normalsize{$HHG_{2,3}$ Phase}]{\includegraphics[scale = 0.11]{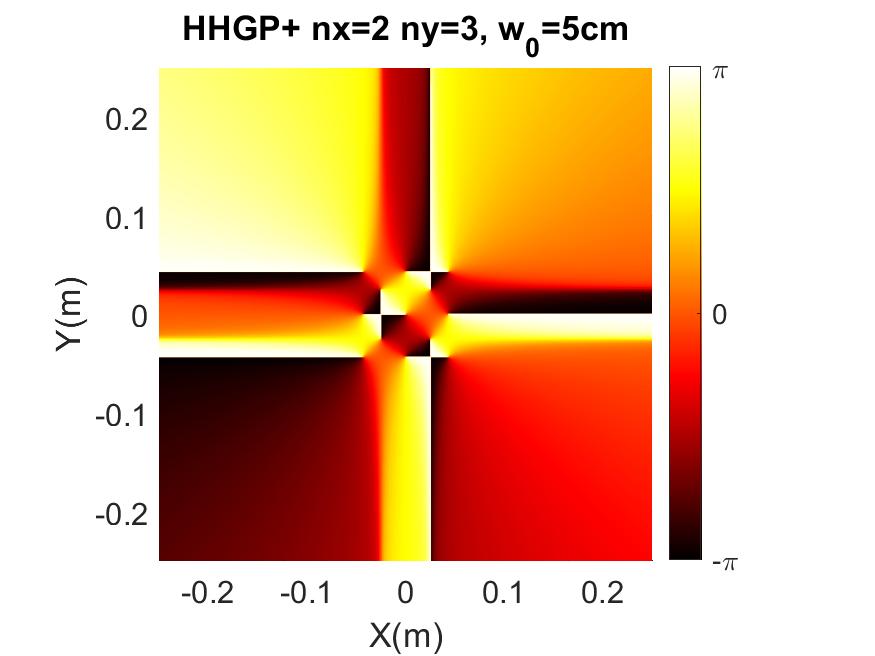}}
	\caption{Evolution of the ellipticity parameter for $HIG_{5,3}$, intensity (a) - e)) and phase (f) - j)) profiles. The chosen values of ellipticity are $\varepsilon =0$, $\varepsilon =1$, $\varepsilon =4.2$ and $\varepsilon =0$ from left to right for a) - d) and d) - h). The waist of the beams is $w_{0} = 5cm$.}
	\label{HelicalEvolutionbegin}
	\end{center}
\end{figure}

\noindent
For these helical Ince-Gauss beams we  identify phase discontinuities at various points of the beam profile , each with value $\pm 2\pi$ (or OAM $\hbar$ per photon) \textit{at} the position of the intensity vortexes, differing from helical Laguerre-Gauss modes \textit{(HLG)}, that exhibit a phase discontinuity of value $\pm 2\pi l$ at the propagation axis of the beam, giving place to a topological charge of $l$ per photon. Examples of these HIG modes are shown in Figure \ref{HelicalEvolutionbegin}, were we identify the phase discontinuities and zero intensity regions of each beam. The initial OAM value of the $HLG$ mode, contained in the optical axis phase indetermination, breaks down into individual $2\pi$ outside said axis, all giving place to zero-intensity regions.\\

While the OAM of a certain HIG photon mode at each intensity vortex is $\hbar$, as a whole, it has been demonstrated that the photon expectation value for the beam can be found to be \cite{Plick}:

\begin{equation}
    \langle \hat{L}_{z}\rangle_{HIG} = \pm\sum_{n,l} \hbar l D_{n,l}^{e}D_{n,l}^{o}\textrm{.}
\end{equation}

\noindent
where $\hat{L}_{z}$ is the OAM operator and $D_{n,l}^{e}$, $D_{n,l}^{o}$ are the decomposition constants of a certain HIG mode transformed into a superposition of even and odd LG modes. The OAM expectation value of non azimuthally symmetric modes is not necessarily an integer number, and is  not a direct average of the modes comprising the LG decomposition. This is because OAM arises from the gradient of the transverse phase structure of the beam. As the argument (phase of a mode) is not a linear function, an OAM value for HIG modes can not be reconstructed from weighted averages of well known HLG modes OAM values. Also, as the decomposition values of the transformation $D_{n,l}^{\sigma}$ are a function of the ellipticity of the modes, $\langle \hat{L}_{z}\rangle$ is also a function of ellipticity, which gives place to HIG modes with different values of $\varepsilon$, $p$ and $m$ (and thus projected into a different basis) but with the same value of OAM.\\

What is actually meant by the ``orbital angular momentum'' of a light field in many cases becomes complicated, due to the fact that there are several definitions, and that the OAM is not an independent quantity of a photon, but instead exists relative to a particular measurement direction and choice of beam axis. For example, a detector that is aligned with an incoming LG beam and whose aperture captures all light will measure the $l$ index of the beam. However if the detector is displaced from the beam axis by some amount the measured OAM will decrease as the light field no longer ``orbits'' the optical axis of the detector. Furthermore, for the case of IG modes, imagine there is a detector which captures only a fraction of the light field around one of several vortices. The detector will register only the topological charge around that vortex, missing the others \--- giving one OAM value. But then if the detector is expanded to include the whole beam another OAM value will be measured. The particulars are beyond the scope of this paper, but for the case of IG modes a very-detailed discussion can be found in Ref.\cite{Plick} of what the OAM ``means'' for IG modes. \\

Since we are interested mostly in applications to quantum communication in this work the most salient measure of OAM is the projection of the state onto some given target vector \--- which is the overlap integral we investigate. In the lab OAM mode sorters, as well as SLMs and single-mode fibers, can implement these mathematical operations physically at the moment of detection.\\

As OAM is an infinite basis that can been used for the transport of information, and the ellipticity parameter of helical vortex beam, as well as the order and degree give place to paraxial beams that carry non-integer quantities of OAM (differing from the usually used Laguerre-Gaussian modes), HIG arise as alternative candidates for optical communications in free space as they provide an extra degree of freedom for their tuning in the $\varepsilon$ parameter given certain $p$ and $m$ that could be taken advantage of given an adequate detection frame, for applications such as classical cryptography and quantum key distribution.

\section{Numerical modeling method}

We model atmospheric turbulence on propagating IG beams by applying a matrix method. First, we define the beam and propagation parameters, such as the matrix size $N$, the matrix physical size $L$ (in microns), the wavelength $\lambda$, beam waist $w_{0}$, order $p$, degree $m$ and ellipticity $\varepsilon$. In order to calculate the mathematical expression for the desired IG field, we decompose the desired IG mode into a superposition of LG modes and insert the resulting mathematical expression into a complex matrix of dimensions $N\times N$. The decomposition is performed because it is easier to implement the mathematical expressions of Laguerre polynomials, compared to Ince polynomials. After that, we select the total distance of propagation $ZT$ and the total number of distance divisions $n_{div}$, in order to evaluate the dynamics of the modes at selected different distances. Next, we choose the atmospheric turbulence parameters, such as the refractive index structure parameter $C_{n}^{2}$, the power spectral density $\Phi_{n}$ and the distance between random phase screens $d_{scr}$, which represent the atmospheric turbulence conditions. Finally, we establish an adequate number of times to propagate the mode $n_{prop}$, in order to get a reliable average result of measured quantities. We do this because of the random nature of refractive index fluctuations in the atmosphere, that result in no beam propagation being the same and thus we have to rely on average results.\\

We use for this propagation procedure, the modified von Karman power spectrum, as it is an extension of the usually used Kolmogorov power spectrum:

\begin{multline}
    \Phi_{n}^{vK}(\kappa) = 0.033C_{n}^{2}\kappa^{-11/3}\frac{\textrm{exp}(-\kappa^{2}/k_{m}^{2})}{(\kappa^{2}+k_{0}^{2})^{11/6}}\quad \\\textrm{for}\quad 0\leq\kappa<\infty\textrm{,} 
    \label{vkarman}
\end{multline}

\noindent
where $\kappa$ is the angular spatial frequency vector, $k_{m} = 5.92/l_{0}$  and $k_{0} = 2\pi/L_{0}$, and $L_{0}$ and $l_{0}$ are the as the outer and inner scales, respectively. The outer scale represents the size of turbulent cells that cause a beam to be randomly deflected from its path, causing fluctuations in the direction and position of the beam at the receiver aperture. The inner scale cells are responsible of the small scale effects that distort the wavefront, resulting in a randomly aberrated phase (and intensity distribution) at the receiver end. 
Since the von Karman model is defined over the entire spectral range $0\leq \kappa < \infty$, it has the advantage 
over the Kolmogorov turbulence model of taking values outside the inner scale. For values inside the inertial subrange $\kappa_{0}<<\kappa<<\kappa_{m}$ both models coincide.\\

Once we define the initial conditions and thus construct the initial beam profile matrix, we implement the following algorithm:
 First, we generate random phase screen of the same size and dimension as that of the beam that represents the atmospheric turbulence. This phase screen is multiplied by the beam profile matrix. Next, we Fourier transform the resulting matrix and propagate it a distance $d_{\mathrm{screen}}$ using a transfer function, to eventually apply an inverse Fourier transform to the result. By doing this we get a new beam profile matrix that has propagates through simulated atmosphere. We repeat this process until a distance corresponding to a multiple of $ZT/n_{\mathrm{div}}$ is reached and at that point, we perform measurements on the scintillation index, overlap and Strehl ratio, by comparing the turbulence propagated beam profile to one with the same initial conditions, but propagated the same distance without atmospheric turbulence ($C_{n}^{2} = 0$). We store the results and repeat the process until the total distance $ZT$ is reached, at which point, we verify the number of propagations performed and if it is not equal to $n_{\mathrm{prop}}$, we perform the whole process again. We do this until reaching the propagation number $n_{\mathrm{prop}}$, and at each distance multiple of $ZT/n_{\mathrm{div}}$ we store the beam intensity and phase transverse profiles. At the end of the cycle, we calculate an average for each of the measurements over all propagations $n_{\mathrm{prop}}$.\\ 

We do the generation of the random phase screens based on works such as \cite{cuatroseis}: We generate a pseudorandom $NxN$ array of complex numbers, and multiply it by $2\pi/(N \Delta_{k})\sqrt{2\pi k^{2}d_{\mathrm{screen}}\Phi_{n}}$, where $\Delta_{k}$ is the spatial sampling interval in the Fourier space. We inverse Fourier transform the result, and get a real space complex random phase field. Then we choose the imaginary part of said phase field as a random phase screen.\\

We use this algorithm to obtain the scintillation index, overlap value and a redefined version of the Strehl ratio, all of which we define below:\\

 The scintillation index $\sigma^{2}$ is a number that represents the intensity variance of a light beam, meaning, it is a number that measures how much the \textit{brightness} of the beam changes during propagation at a certain point (scintillation). The scintillation index is given by \cite{tata}:
\begin{equation}
    \sigma^{2}(z') = \frac{\langle I^{2}(z')\rangle-\langle I(z')\rangle^{2}}{\langle I(z')\rangle^{2}} = \frac{\langle I^{2}(z')\rangle}{\langle I(z')\rangle^{2}} -1\textrm{,}
\end{equation}

\noindent
where $I(z')$ is the intensity of the beam profile at a certain distance of propagation $z'$. The smaller the value of $\sigma^{2}$ the more \textit{stable} the beam is.\\

The Strehl ratio $SR$ is arbitrarily defined, in the case of this work as:
\begin{equation}
    SR(z') = \frac{I(z')}{I_{0}(z')}\textrm{,}
\end{equation}

\noindent
where $I_{0}$ is the total intensity of the beam while propagating with no turbulence. This quantity measures how much of the turbulence propagated beam intensity goes out of a window defined by the size of the beam without turbulence. The closer the value of the SR is to 1, the less the beam is distorted or deflected by atmospheric turbulence.\\

Finally, the overlap \textit{OV} (also refered as fidelity) of a light beam refers to how much the initial beam power is kept in the original selected mode after propagation, and is defined by a ratio of inner products of the original beam and the propagated one \cite{forbes}:

\begin{equation}
    OV(z') = \frac{\int\int_{-\infty}^{\infty}E(x,y,z')\bar{E'}(x,y,z')dxdy}{\int\int_{-\infty}^{\infty}E(x,y,z')\bar{E}(x,y,z')dxdy}\textrm{,}
    \label{overlap}
\end{equation}

\noindent
where $E(x,y,z')$ refers to the transverse electrical field of the mode propagated a distance $z'$ \textit{without} turbulence and $E'(x,y,z')$ the mode after propagation through turbulent media, $\bar{E}(x,y,z')$ and $\bar{E'}(x,y,z')$ are the respective complex conjugates of the fields. Assuming normalized electric fields, $OV$ can take values between $0$ and $1$, meaning total lost or total conservation of the original beam transverse profile.\\

\begin{figure*}[h]
	\begin{center}
	\subfloat[Reference window]{\includegraphics[scale = 0.15]{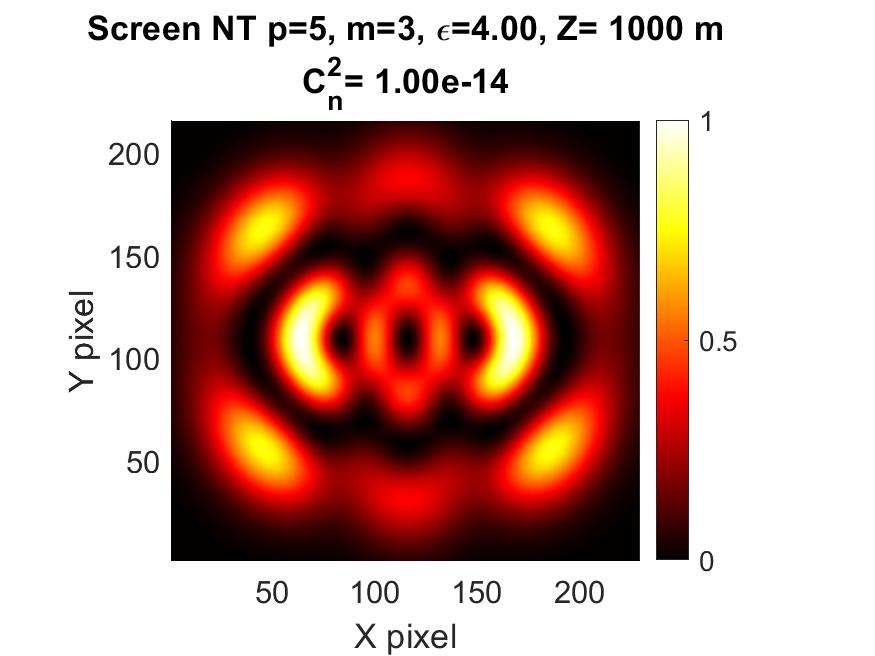}}
	\subfloat[Turbulence beam window]{\includegraphics[scale = 0.15]{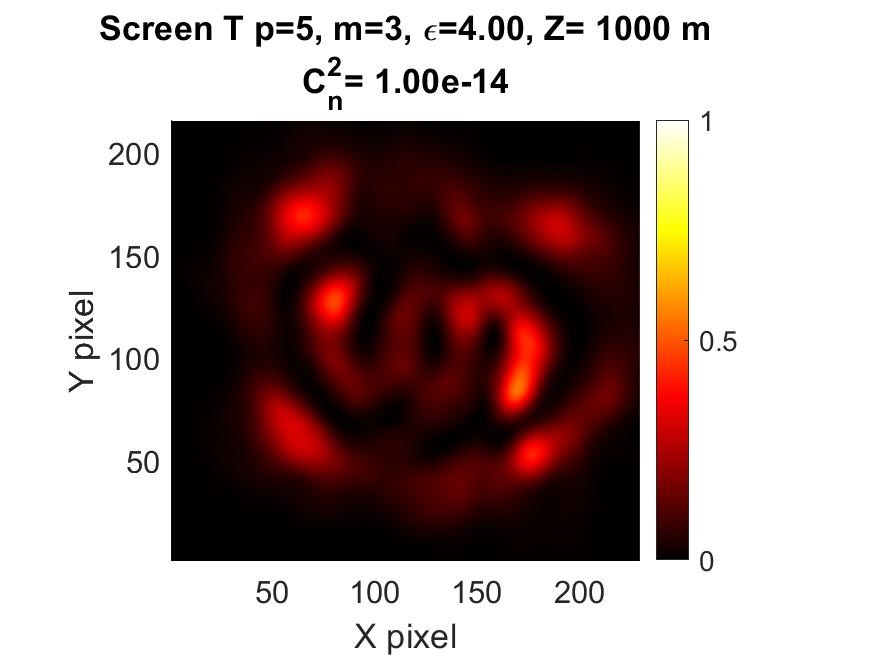}}\\
	\caption{Example of a) reference window a) and b) measurement window for $HIG_{5,3}^{4}$. As can be graphically seen, some of the power from the turbulence propagated beam escapes from the measurement window, heavily affecting the value of the Strehl ratio, the Scintillation index is affected by the distortion on the intensity profile of the beam, caused by the turbulence.}
	\label{region}
	\end{center}
\end{figure*}

The SI and the SR are very shape-dependent measurements. For the SI, this is because the mean value of intensity varies heavily from point to point depending on the geometrical shape of the transverse intensity profile. For the SR, it is because the size of the complete beam profile matrices is in general bigger than that of the transverse profile itself. Therefore, we delimit a reference region for the evaluation of these values. We define said limit by using the last elements of the no-turbulence propagated beam profile that have an intensity greater than or equal to $1/e^{2}$ the maximum intensity value of the whole beam profile. We picture this limit region for measurements in Figure \ref{region}, where we show a $HIG_{5,3}^{2}$ mode, propagated with and without turbulence. The no-turbulence propagated profiles define the SR and SI measurement region for the turbulence propagated beam characterization.\\

We investigate robustness of helical vortex beams by propagating beams with varying $\varepsilon$, $p$ and $m$, putting special attention to the overlap measurement, as it is an indicator on how much information of a beam is kept in the initial transverse propagation mode: All the other propagation parameters are the following: $N = 1024$, $L = 50 cm$, $\lambda = 632.8 nm$, $w_{0} = 10000 \mu m$, $l_{0} = 1 cm$, $L_{0} = 3 m$, $n_{\mathrm{prop}} = 200$, $d_{\mathrm{screen}} = 20 m$, $n_{\mathrm{div}} = 20$. $C_{n}^{2}$ varied from $C_{n}^{2} = 10^{-14}m^{2/3}$ to $C_{n}^{2} = 10^{-16}m^{2/3}$, representing strong to weak turbulence values for the chosen wavelength \cite{Andrews}. We choose these arbitrarily in order to both perform more accurate simulations without compromising too much computation time due to the raw number modes to propagate, and to easily identify the effects that the choose of beam parameters such as $p$, $m$ and $\varepsilon$ have on the simulated performance of the beam. It is worth mentioning that the simulations mainly study the properties of the beam itself and not on a particular FSO communication system. The general characteristics found for HIG beams hold for other choose of parameters which may depend on a particular FSO communication system.

\section{Propagation simulation results and analysis}

\subsection{Propagation of Helical Ince-Gaussian modes varying the ellipticity parameter}

\begin{figure*}[h]
	\begin{center}
	\subfloat[\normalsize{$0 m$}]{\includegraphics[scale = 0.15]{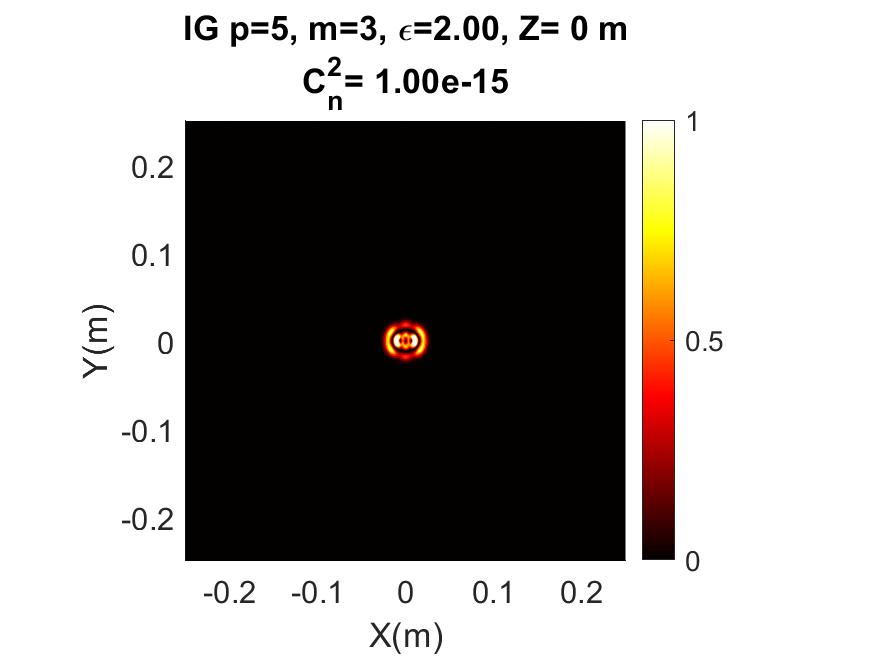}}
	\subfloat[\normalsize{$500 m$}]{\includegraphics[scale = 0.15]{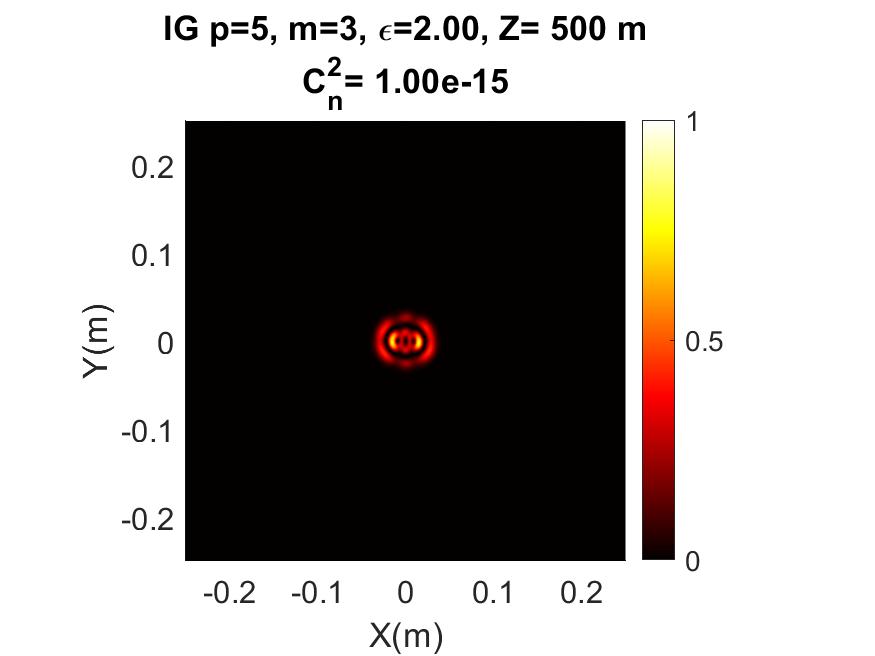}}
	\subfloat[\normalsize{$1000m$}]{\includegraphics[scale = 0.15]{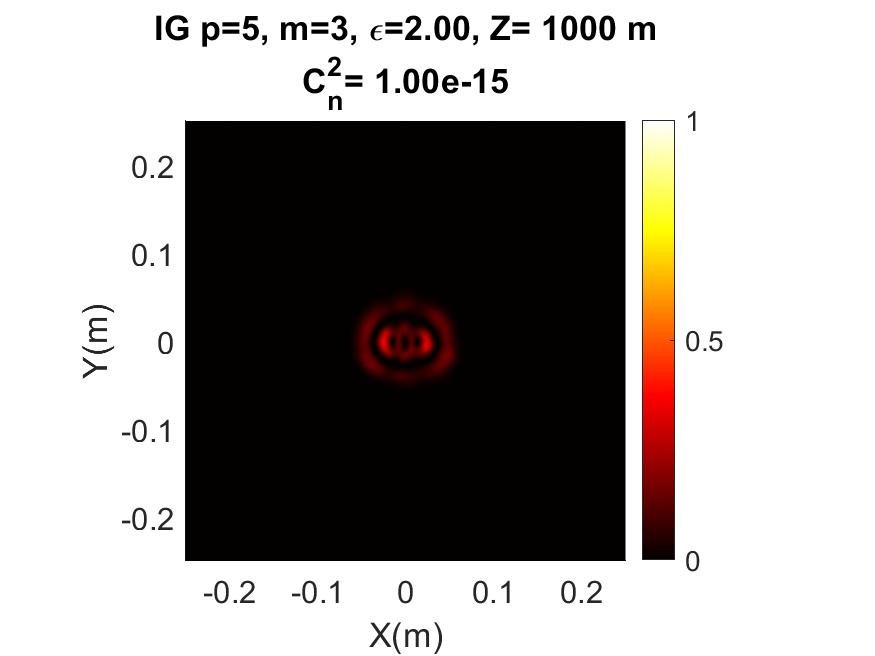}}\\
	\subfloat[\normalsize{$1500m$}]{\includegraphics[scale = 0.15]{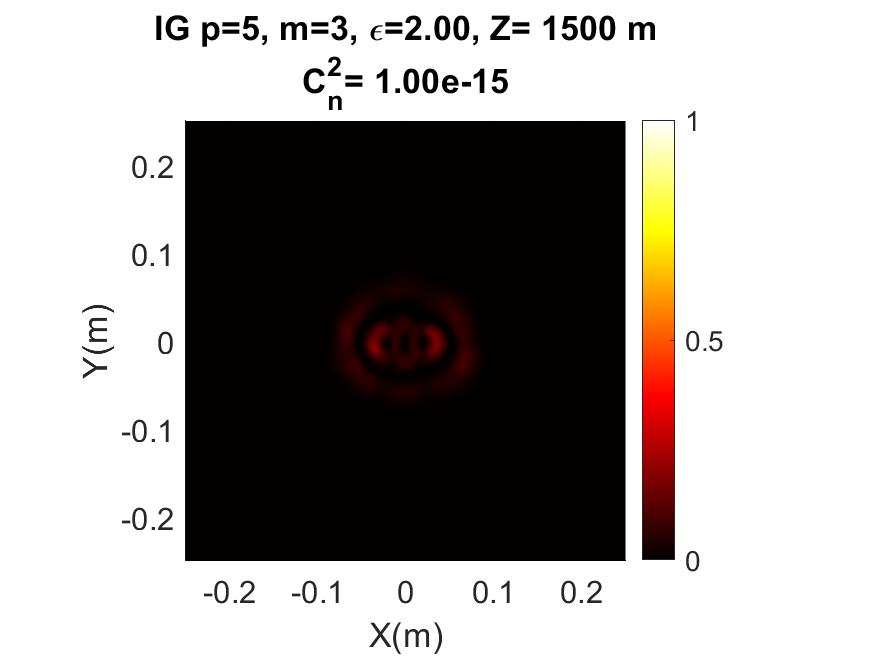}}
	\subfloat[\normalsize{$2000m$}]{\includegraphics[scale = 0.15]{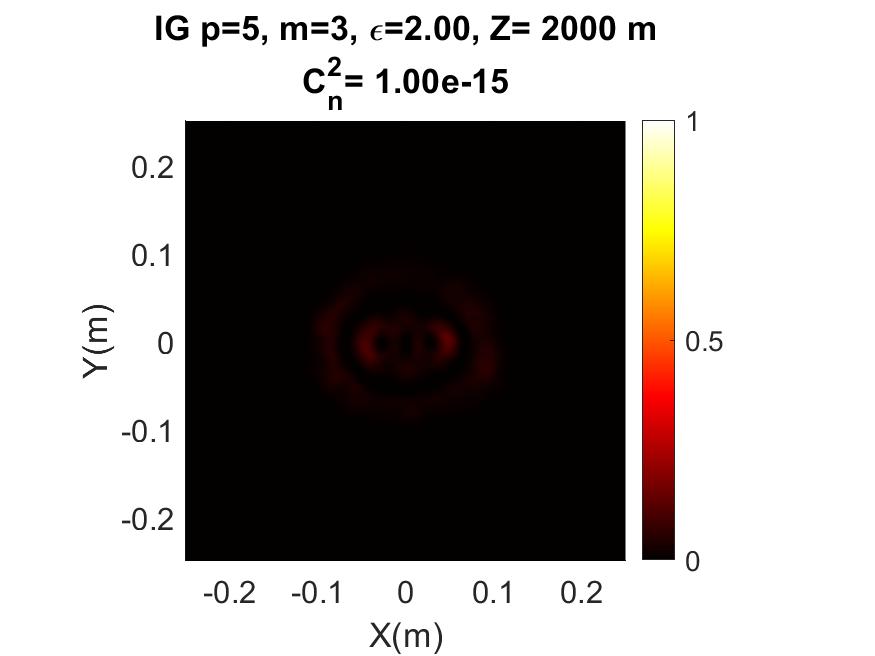}}
	\caption{Propagation of a $HIG_{5,3}^{2}$ mode through 2km of turbulent atmosphere with $C_{n}^{2}=10^{-15} m^{2/3}$.}
	\label{turbulentp}
	\end{center}
\end{figure*}

\begin{figure}[h]
    \centering
    \subfloat[\normalsize{Overlap value}]{\includegraphics[scale = 0.20]{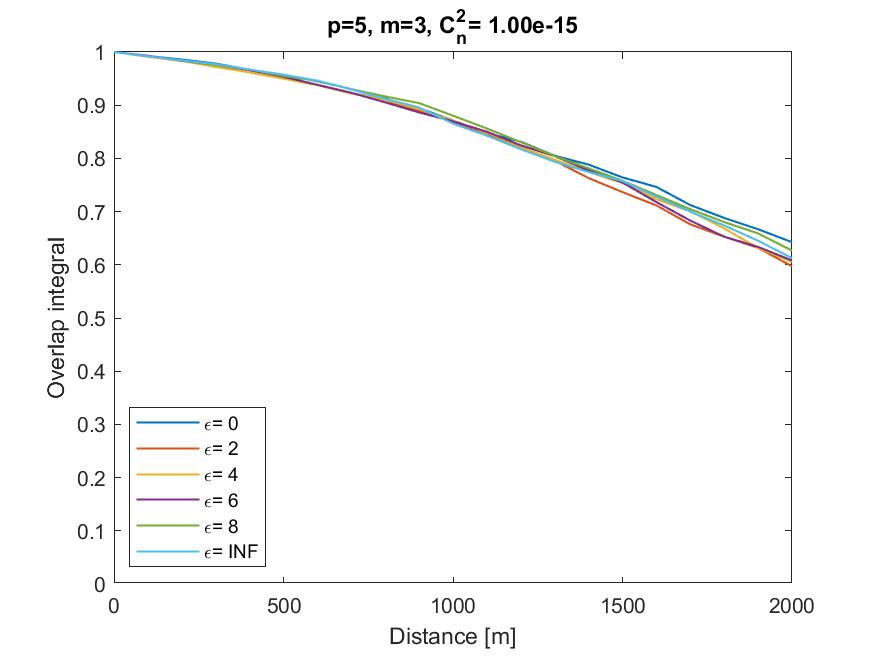}}\\
    \subfloat[\normalsize{Scintillation index}]{\includegraphics[scale = 0.20]{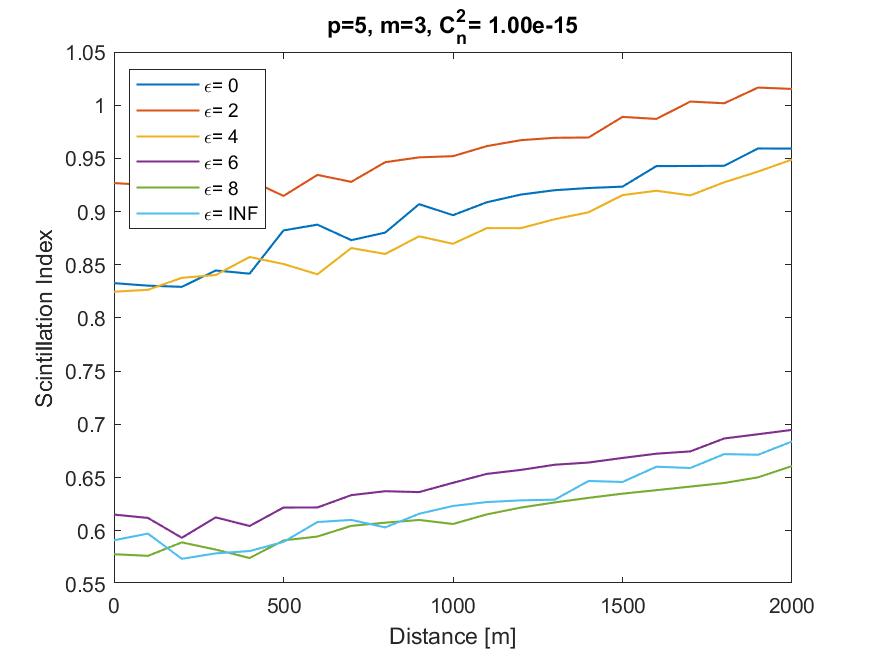}}\\
    \subfloat[\normalsize{Strehl Ratio}]{\includegraphics[scale = 0.20]{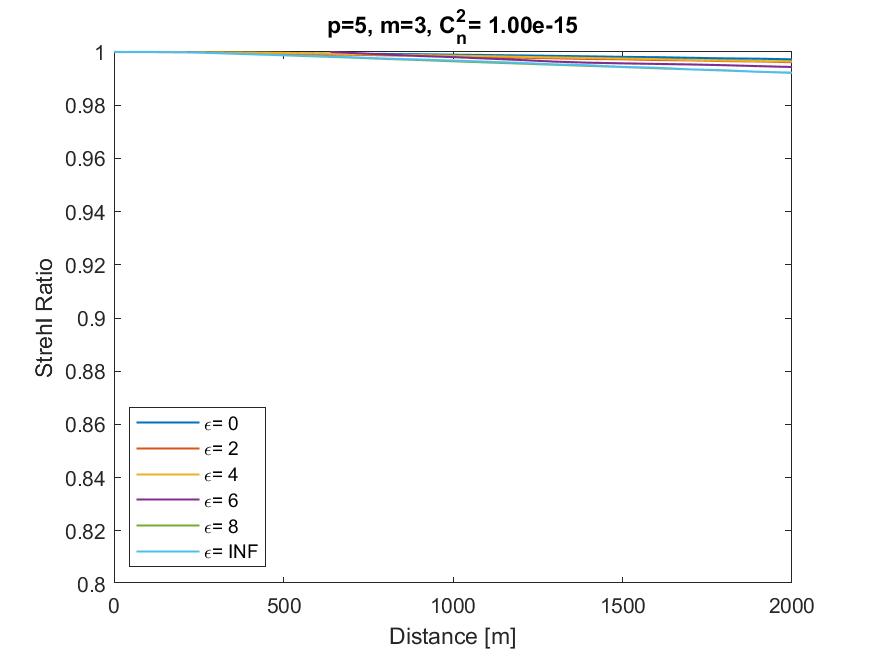}}
    \caption{Propagation characterization of several $HIG_{5,3}$ modes with varying ellipticity values at $C_{n}^{2} = 10^{-15} m^{-2/3}$.}
    \label{53 ellip measu}
\end{figure}

We choose a $HIG_{5,3}$ mode for the variation on the ellipticity parameter of $HIG$ modes through atmosphere, as well as a refractive index structure parameter of $C_{n}^{2}=10^{-15}m^{2/3}$ for a $2km$ propagation. The ellipticity parameter values were chosen so that a change of the beam from $HLG$ to $HHG$ modes could be appreciated. We present in Figure \ref{turbulentp} an example of the propagation through turbulent atmosphere of a $HIG_{5,3}^{4}$ mode, and the results of this simulations for the SI, SR and OV in Figure \ref{53 ellip measu}. The results are shown in terms of distance and turbulence strength instead of the Fried parameter $r_{0}$ to emphasize  \\

The ellipticity parameter appears to have a strong impact in the SI of the modes, with certain $\varepsilon$ values having a considerably bigger value than others. This can be attributed to the geometrical structure nature of the measurement, that takes into account the intensity mean values for all points of the mode. By having a mode in which the intensity of the beam is mainly concentrated at a certain region of the transverse profile, as happens for $HIG_{5,3}$ with low $\varepsilon$, the difference in intensity between the outer parts of the mode and the inner parts, that are almost zero, make the scintillation index much bigger. In contrast, higher values of $\varepsilon$ make the intensity of the beam more evenly distributed, and thus, seem to help diminish the SI value. Regarding the SR value, $\varepsilon$ seems to have a very subtle effect that is not very appreciable with a $C_{n}^{2} = 10^{-15} m^{-2/3}$ refractive index structure parameter.\\

However, the most important result of the simulation is the fact that the OV of the modes as a function of distance is almost the same for all values of $\varepsilon$ at all distances, with no particular preference for a value that would make the beam profile more resilient to mode leakage by turbulence effects. This result implies that for a certain mode, defined by the order $p$ and degree $m$, any family of modes ($\varepsilon$) that is chosen to propagate this mode through atmosphere would have around, if not the same, effectiveness. Thus, the robustness of a beam through the atmosphere can be consider to be effectively independent of the chosen symmetry or $\varepsilon$. In order to corroborate this result, we execute more simulations.\\

\begin{figure}[h]
    \centering
    \subfloat[\normalsize{Overlap for strong turbulence}]{\includegraphics[scale = 0.20]{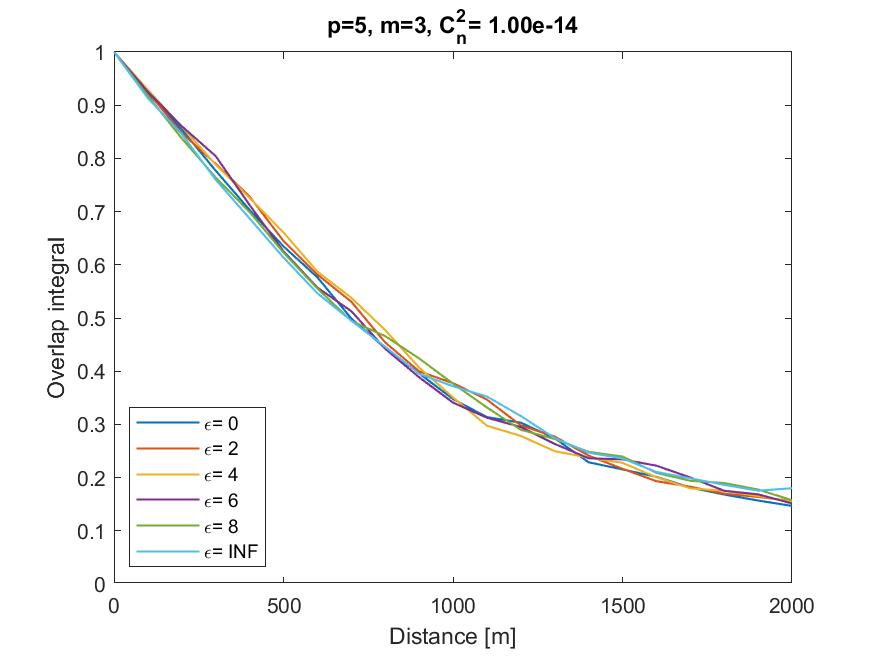}}\\
    \subfloat[\normalsize{Overlap for medium turbulence}]{\includegraphics[scale = 0.20]{fig5a.jpg}}\\
    \subfloat[\normalsize{Overlap for weak turbulence}]{\includegraphics[scale = 0.20]{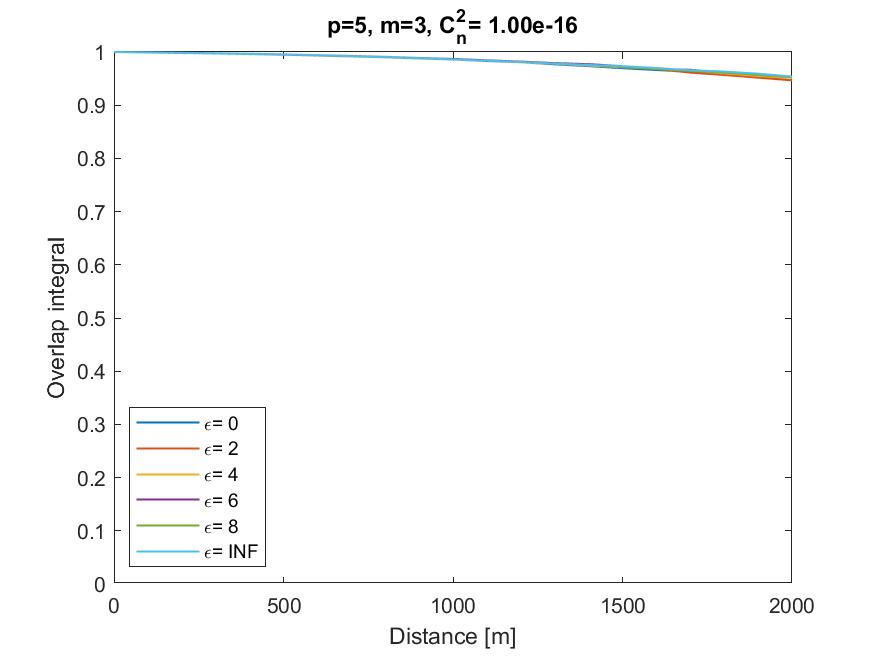}}
    \caption{\centering \normalsize{Comparison of overlap value obtained for $HIG_{5,3}$ modes with varying ellipticity for different turbulence strength}}
    \label{ellip turb comparison}
\end{figure}

\begin{figure}
    \centering
    \subfloat[\normalsize{Overlap value}]{\includegraphics[scale = 0.20]{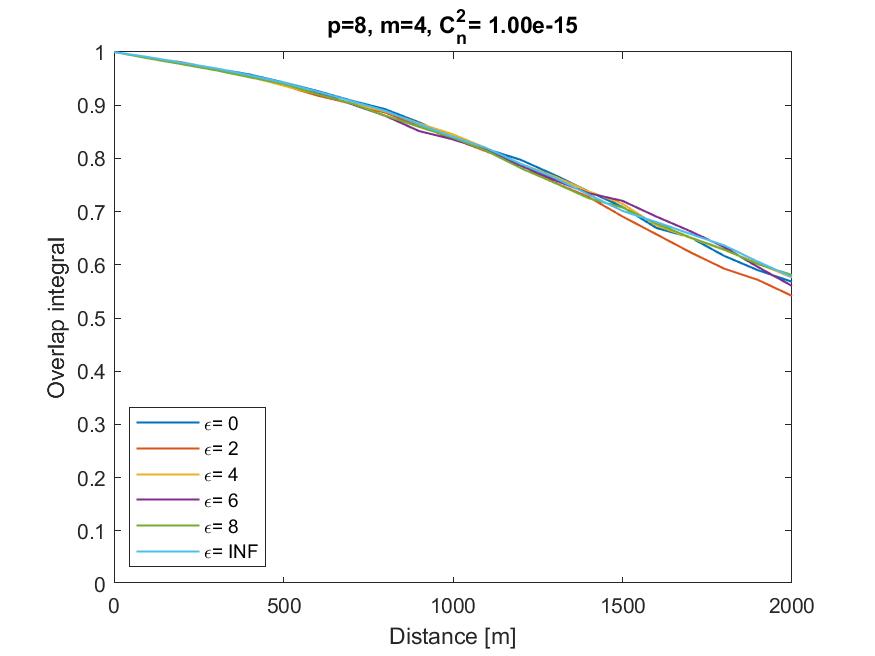}}\\
    \subfloat[\normalsize{Scintillation index}]{\includegraphics[scale = 0.20]{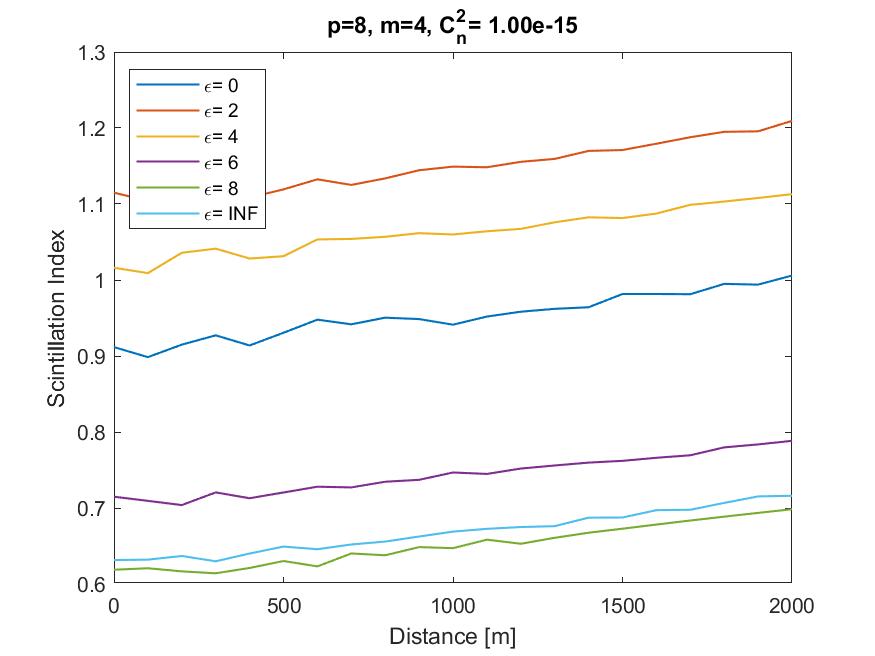}}\\
    \subfloat[\normalsize{Strehl Ratio}]{\includegraphics[scale = 0.20]{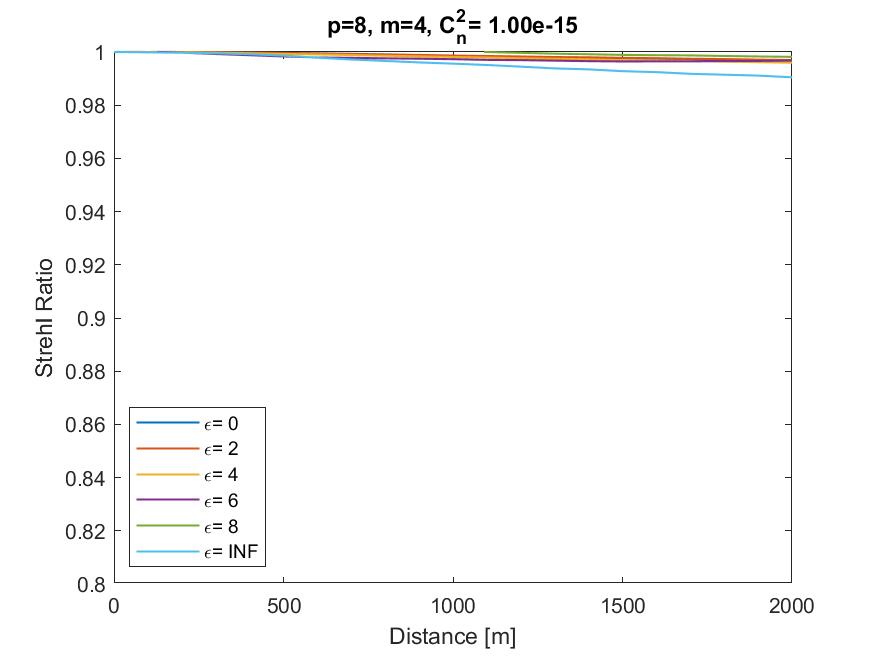}}
    \caption{\centering \normalsize{Propagation characterization several $HIG_{8,4}$ with varying ellipticity values at $C_{n}^{2} = 10^{-15} m^{-2/3}$, showing that the OV holds for different $\varepsilon$ even with different modes.}}
    \label{82measu}
\end{figure}

First, we propagate the same selected ellipticity modes through weak and strong turbulence and present a comparison of the measurement of the OV for these conditions in Figure \ref{ellip turb comparison}. We also perform a propagation simulation for a $HIG_{8,4}$ mode with varying $\varepsilon$, to corroborate if the observed characteristics of the propagated beams with $\varepsilon$ are consistent for other sub-basis of modes. The results of these propagations are shown in Figure \ref{82measu}.\\

The results from Figures \ref{ellip turb comparison} and \ref{82measu} show that the general characteristics of the beam propagation discussed for $HIG_{5,3}$ for varying $\varepsilon$ hold for both different turbulence strengths and different modes. The relation of the OV of the propagated beam with its $\varepsilon$ holds for all cases, which indicates that there are other factors that influence performance of helical OAM-carrying beams. However, this also means that depending in applications such as cryptography, any $\varepsilon$ value (which is continuous quantity) could be use and the robustness of the beam would be constant.

\subsection{Propagation of Helical Ince-Gaussian modes, varying the degree ($m$) parameter}

The next propagation simulation  involved the the performance of the beam depending on the degree parameter ($m$) for $HIG$ modes, recalling that this parameter is closely related to the OAM of helical beams, specially when the ellipticity of the mode is $\varepsilon =0$. In this case, the $HIG_{p,m}^{0}$ transforms into a $HLG_{n,l}$ mode with $l=m$ and $OAM = l\hbar$ per photon.

\begin{figure}
    \centering
    \subfloat[\normalsize{Overlap value}]{\includegraphics[scale = 0.20]{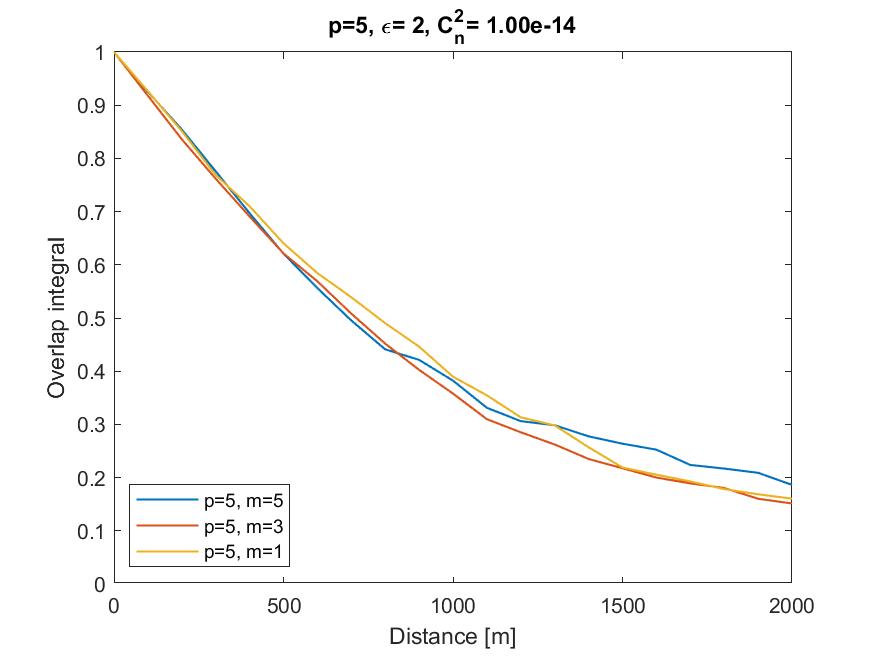}}\\
    \subfloat[\normalsize{Scintillation index}]{\includegraphics[scale = 0.20]{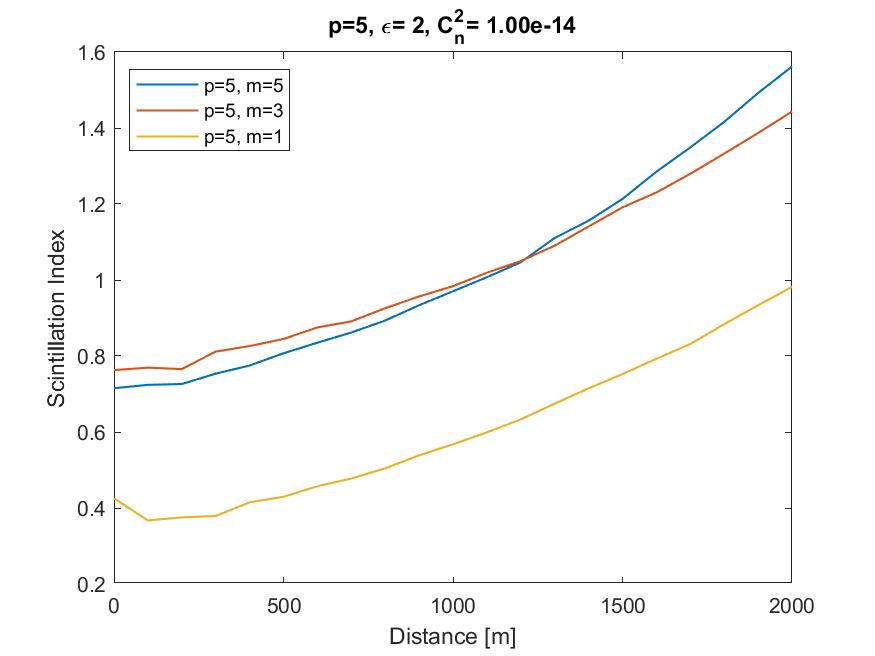}}\\
    \subfloat[\normalsize{Strehl Ratio}]{\includegraphics[scale = 0.20]{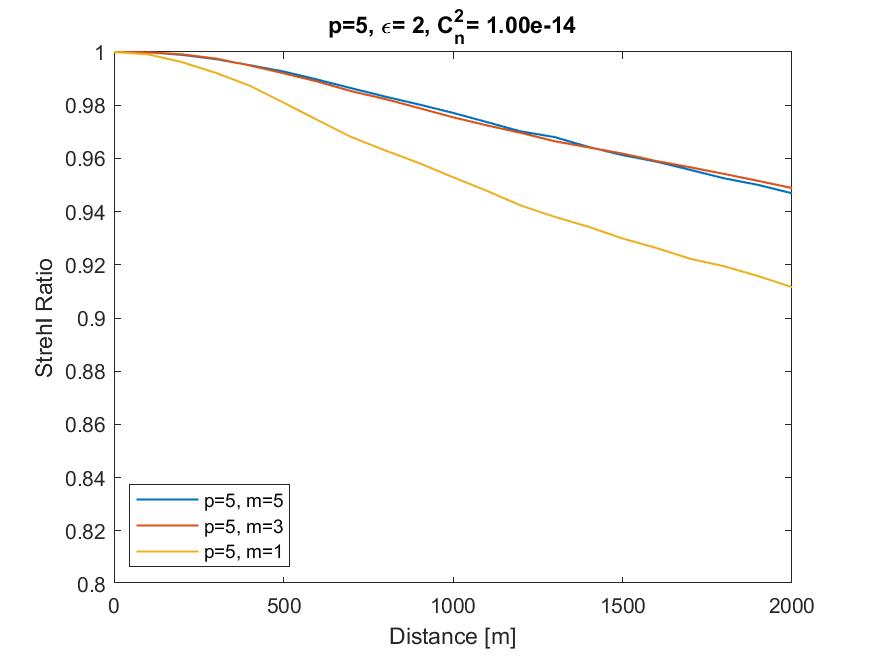}}
    \caption{Propagation characterization of several $HIG_{5,m}^{2}$ with at $C_{n}^{2} = 10^{-14} m^{-2/3}$ by varying the degree of the mode.}
    \label{pfixedcar}
\end{figure}

The simulation performed involved $HIG$ modes with fixed order $p=5$ and a turbulence strength of $C_{n}^{2} = 10^{-14}m^{-2/3}$. The results on the characterization of the propagation are shown in Figure \ref{pfixedcar}.\\

 It is interesting to notice that for the SI and  SR at some points of the propagation the preferred $m$ changes, but in general higher values of $m$ behave better. This can be explained by noticing that lower values of $m$, or rather, bigger differences in $p-m$ generate more elongated modes than those with lower $p-m$ value. Also, the window in which these quantities are initially checked is more elongated in the $x$ (or $y$) axis and the turbulence effects on the beams tend to distort it and deviate beam intensity from the measurement window. Regarding the OV, we notice that for longer propagations, the turbulence strength has a bigger effect on the leakage of power into other modes from the initial beam: the bigger values of $m$ performed better, which is contrary to the initial assumption on the performance of the beam with the $m$ parameter based in the comparison to $LG$ modes. Because of this, it is congruent to assume that not only the $m$ parameter, but the order ($p$) parameter influences the effectiveness of these beams.\\

\subsection{Propagation of Helical Ince-Gaussian modes, varying the order ($p$) parameter}

\begin{figure}
    \centering
    \subfloat[\normalsize{Overlap value}]{\includegraphics[scale = 0.20]{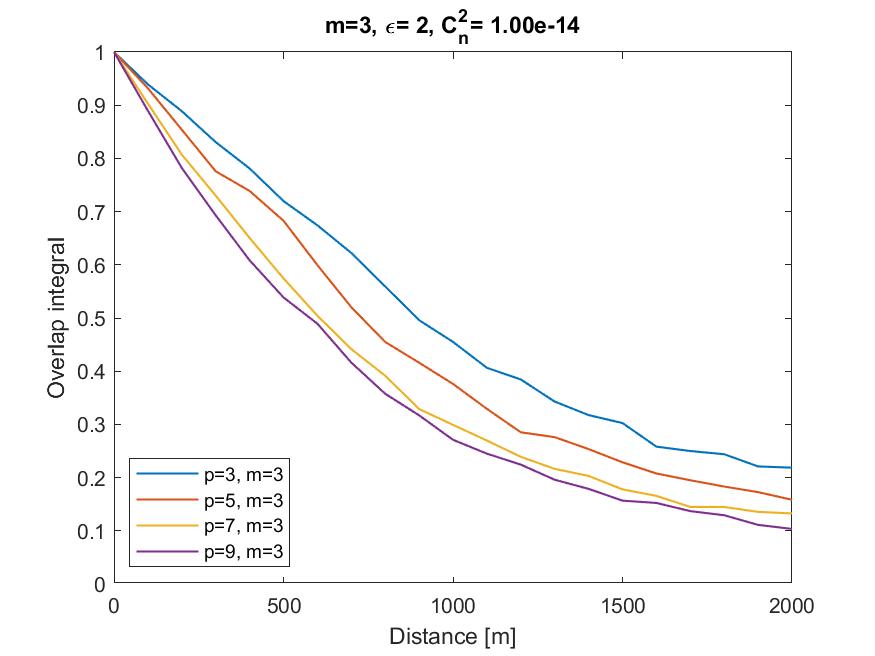}}\\
    \subfloat[\normalsize{Scintillation index}]{\includegraphics[scale = 0.20]{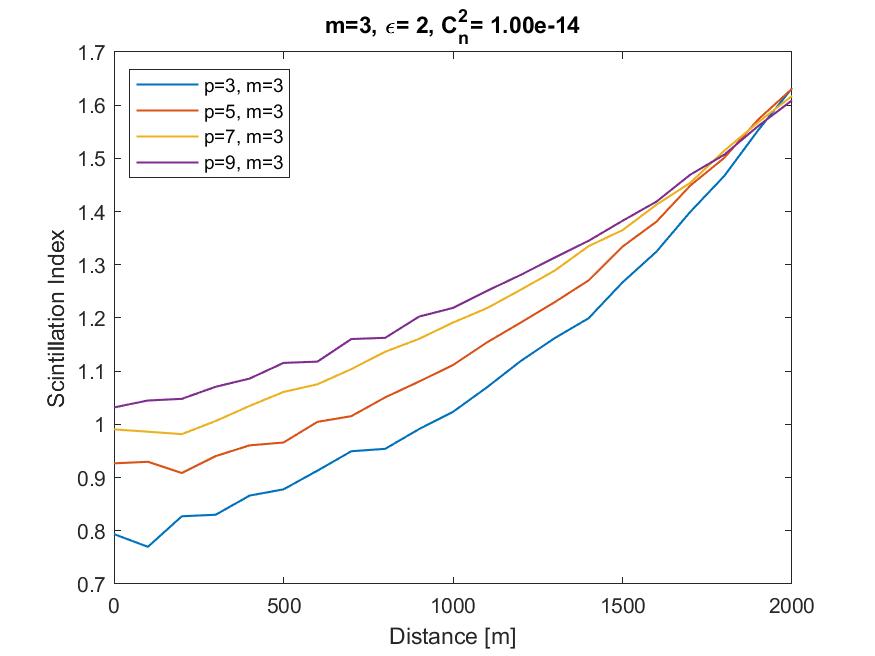}}\\
    \subfloat[\normalsize{Strehl Ratio}]{\includegraphics[scale = 0.20]{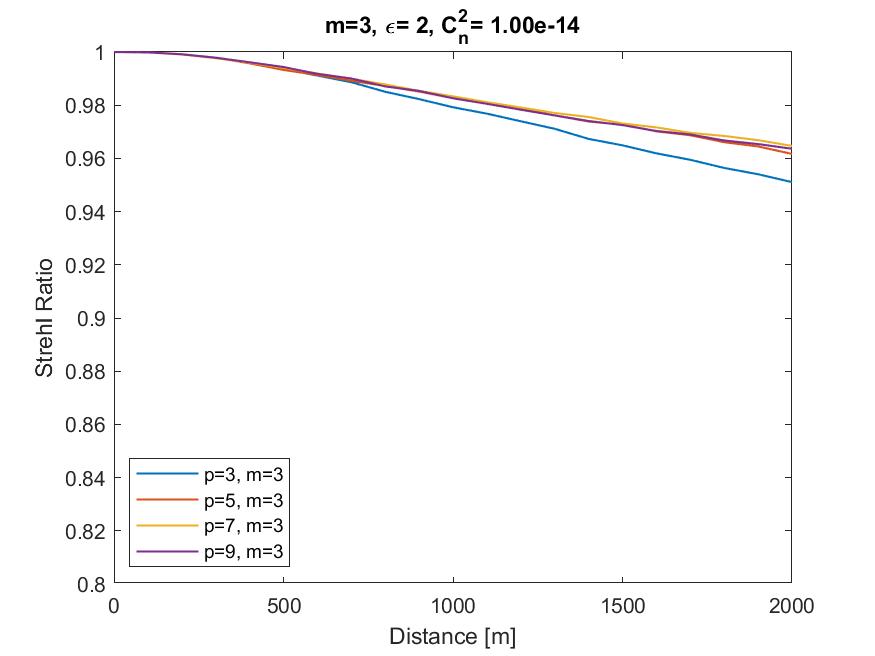}}
    \caption{\centering \normalsize{Propagation characterization of several $HIG_{p,3}^{2}$ with at $C_{n}^{2} = 10^{-14} m^{-2/3}$ by varying the order of the mode.}}
    \label{mfixedcar}
\end{figure}

The effects of the order $p$ parameter on $HIG$ beams are studied in order to, in conjunction with the results obtained from varying the degree $m$ parameter, obtain information about how the physical structure of the $HIG$ beams affects their performance in turbulent media. We perform the simulation of propagation for a  $HIG_{p,3}^{2}$ mode with a refractive index structure constant of $C_{n}^{2}= 10^{-14}m^{-2/3}$ for varying values of $p$. The results of the simulation are shown in Figure \ref{mfixedcar} for the characterization measurements of the propagation.\\

We obtain a similar result to the one of propagation of modes with a fixed $m$ parameter. For these modes a smaller $p$ value is preferred for the SI up to a certain distance, where most of the modes seem to converge to a very similar value except for the lower $p=3$ mode. Regarding the Strehl Ratio, there is no clear mode with radically better performance. In general, a bigger impact on these two measurements (due to their effect in the transverse geometry of the mode) can be attributed to the degree $m$ of a mode, rather than to the order $p$.\\

However, for the OV measurement, there is a clear preference in performance for lower $p$ modes, that is even more noticeable than the preference for higher values of $m$, having the $HIG_{3,3}^{2}$ around double the overlap value compared to the worst performing mode $HIG_{9,3}^{2}$.\\

The results for the overlap for both fixed $p$ and $m$ parameters show a remarkable condition. Rather than just looking at these parameters separately, one could look at the results for modes with different $p-m$ value. In both Figures \ref{pfixedcar} and \ref{mfixedcar}, the modes with an smaller difference value perform better than the others. To better understand this result, one could look at the relation of this difference with the radial parameter $n$ in $HLG$ modes, where $n=(p-m)/2$. As the radial number $n$ increases, and so does the number of ring phase discontinuities of the transverse mode, the performance of the beam is worse in turbulence. In the case of other values of ellipticity, this difference $p-m$ also determines the number of (elliptical) phase discontinuities rings, which increases with $p-m$ linearly. The more phase discontinuity regions the beam has, the easier it is for turbulence effects to affect the structure of the modes as it demarcates a sharp distinction between regions in the phase profile. For smooth transitions (lower $p-m$ or $n$) one could imagine that errors caused by phase changes to be less stark.  Another way to interpret this result would be that, as the $p$ value has more impact on the robustness of the mode than $m$, and $p$, represents a family of modes that share the same Gouy phase and form a complete sub-basis of orthogonal modes \cite{bandres2}, lower $p$ modes perform better. For lower $p$, there are less modes within the same sub-basis of orthogonal modes in which the mode can leak to, with the different possible values of $m$ being the most probable modes to leak to. However, there may also be leakage into other modes due to loss of coherence.\\

\section{Conclusions and discussions}

We have shown that, remarkably, the ellipticity parameter of the modes $\varepsilon$ does not strongly impact the overlap performance of the modes, and rather is the combination of order $p$ and degree $m$ and their difference $p-m$ the parameters that do. One could expect this to be different because $\varepsilon$ represents a non-discrete degree of freedom that ranges from $0$ to $\infty$. The fact that the difference of performance for each of these discrete values is not prominent is rather striking.  This behavior can be  explained by the fact that $p$ and $m$ define the complexity of a mode and $\varepsilon$ merely defines the basis, or rather, the symmetry system in which the mode is projected. In particular, $p$ defines a sub-family of coherent modes with the same Gouy phase. This means that in a FSO communication system whichever chosen basis (ellipticity) to settle a mode into, the mode should keep the information with barely the same fidelity for a given sub-basis, at least without considering a particular measurement system and for the turbulence model model used in this work.\\

The overlap value can be optimized with the correct $p$ and $m$ values depending on the optical system. All of $p$, $m$, $\varepsilon$ had an impact on the scintillation index and strehl ratio measurements, by giving the beam its geometrical shape and axial symmetry, with less \textit{elongated} modes being preferred. Although the way in which these two parameters were measured in this work was by taking the original beam without propagation as basis for the detection window, in an actual FSO communication system these measurements are heavily dependent on the characteristics of the receivers end and the optical elements used to measure the modes.\\

Works similar to ours have recently been published, like the one from Gu et al \cite{Gu}. In this work, simulations of Helical Ince-Gauss beams were done, paying special attention to analysis of the robustness of the modes as information carriers according to the ellipticity parameter, concluding that it plays an important role on the performance of the modes, as well as the degree $m$ parameter. Another study by Zhu et al was done \cite{Zhu:20} in which the robustness of entangled modes in turbulent atmosphere can be modified by the $\varepsilon$ parameter. Comparing their results to the ones presented in this work, considering both the simulation and methods used, there was a good agreement on the obtained values when converting to the Fried parameter $r_{0}$. However, there was not such agreement on the interpretation of the results. While for these works the effect of ellipticity can be considered significant, for us it is in fact practically negligible compared to the choose of order $p$, based on the overlap values presented. As such, we have demonstrated that the actual choice of modes rather than the basis in which they are constructed affects more heavily the performance of these modes of light as information carriers. This result may seem to be contradicted by works like the one developed by Cox et al \cite{Cox} where it is shown that a subset of Hermite-Gaussian modes is more resilient to atmospheric turbulence than similar Laguerre-Gauss beams (which both represent a different value of ellipticity), by means of using and optical system that simulates turbulence using Spatial Light Modulators. However, there are significant differences in the turbulence models (Kolmogorov and von Karman) utilized in their simulation, the number of modes propagated (which are significantly more in this work) and the experimental means in which the turbulence measurements were obtained, compared to the results of the purely computational simulation done in this work. Here, only the nature of the beams itself is considered and no the experimental setup in which they may be utilized.  The obvious and very important next step to take in regards to the study of OAM carrying beams as information carriers in free space would be to study the behavior of different families of solutions of the PWE beyond the usually studied Laguerre-Gauss, Hermite-Gauss and Bessel-Gauss beams in actual FSO communications.\\

Besides the considered parameters of elliptical vortex beams, it is necessary to add the polarization of light for a more complete result, which results in an additional degree of freedom and could provide other means to optimize the robustness of the modes. Another interesting thing to study more deeply could be the dynamics of the phase vortexes for helical beams: how their distribution changes with atmospheric effects and how this distribution change modifies the orbital angular momentum and information carrying efficiency of the modes. This was done by Lavery \cite{Lavery} but only on the family of Laguerre-Gaussian modes. However, the most important steps to take would be to design and perform experiments in FSO communication systems to test the simulations results and check the feasibility to use these modes to carry information in free space. Nonetheless, the results obtained depicted the nature of the HIG modes through turbulent atmosphere and thus could be used as a reference when designing a specific FSO communication system based on OAM carrying beams.\\

\section*{Funding}
This work was supported by CONACYT, México (grants 293471,293694, Fronteras de la Ciencia 217559).

\section*{Acknowledgments}
We acknowledge support from CONACYT, México.
\section*{References}

\providecommand{\newblock}{}

\end{document}